# Converting a Systems Dynamic Model to an Agent-based model for studying the Bicoid morphogen gradient in *Drosophila* embryo

MARIAM KIRAN, Department of Computer Science, University of Sheffield, Sheffield, UK, m.kiran@dcs.shef.ac.uk

WEI LIU, Cancer Research UK Cambridge Institute, University of Cambridge, Cambridge, UK, wei.liu@cruk.cam.ac.uk

The concentration gradient of the Bicoid morphogen, which is established during the early stages of a *Drosophila melanogaster* embryonic development, determines the differential spatial patterns of gene expression and subsequent cell fate determination. This is mainly achieved by diffusion elicited by the different concentrations of the Bicoid protein in the embryo. Such chemical dynamic progress can be simulated by stochastic models, particularly the Gillespie alogrithm. However, as with various modelling approaches in biology, each technique involves drawing assumptions and reducing the model complexity sometimes limiting the model's capability. This is mainly due to the complexity of the software modelling approaches to construct these models. Agent-based modelling is a technique which is becoming increasingly popular for modelling the behaviour of individual molecules or cells in computational biology. This paper attempts to compare these two popular modelling techniques of stochastic and agent-based modelling to show how the model can be studied in detail using the different approaches. This paper presents how to use these techniques with the advantages and disadvantages of using either of these. Through various comparisons, such as computation complexity and results obtained, we show that although the same model is implemented, both approaches can give varying results. The results of the paper show that the stochastic model is able to give smoother results compared to the agent-based model which may need further analysis at a later stage. We discuss the reasons for these results and how these could be rectified in systems biology research.

## 1. INTRODUCTION

Biological systems can be studied as collections of complex systems ranging from small bacterial interactions to large-scale cell behaviour in tissues. One key feature of these complex systems is their adaptiveness to the changing environmental conditions. This ability to cope and survive, makes these systems extremely robust and are often used as inspirations in building engineering applications. However, studying these systems is a difficult task as they are too complex and intricate to completely understand.

Traditionally numerical equations involving differentiation techniques were used to represent such dynamic systems. Dynamic systems change with time and the conditions of the system. Examples of such equations can be observed in Newton's law of motion or the Navier-Stokes equation for particles and forces. Using a dynamic equation, a system can be represented as a derivative of time (equation 1).

$$X = \frac{\delta x}{\delta t} = f'(x)$$

(eq.1)

The above equation shows how the change in system is represented with time, where *X = (x(1), x(2), ..., x(k))* and *k* is the number of states the system can exist in, usually represented as time periods. The state of a system is given as a property of every element in the system at the moment in time. This includes the properties of the individuals in the system, the environmental conditions and the other attributes involved. Basically, it is a snapshot of the system at time t, which can be taken between the starting time t = 0 till the end of the simulation run. In some systems, it is possible to deduce the description of a system at time *t = t + 1* if the state at *t = t* is known.

Complex systems are emergent systems, which means that it is sometimes difficult to anticipate how the system would behave at the next time step. Also complex systems are referred to as *irreversible* which means that although it is possible to predict future states, it is next to impossible to deduct the past states, even if the present and future states are known.

## 2. THE PROBLEM

Bicoid morphogen gradient establishment, which takes place during early embryo development in *Drosophila melanogaster*, is also a dynamic system allowing the Bicoid molecules to diffuse along the embryo anterior-posterior (A-P) axis in different developmental stages. Such protein concentration gradient is sensed by downstream genes and induces differential spatial pattern of gene expression [Driever and Nüsslein-Volhard 1988a; 1988b; St Johnston et al. 1989; Driever and Nüsslein-Volhard 1989; Struhl et al. 1989; Ephrussi and Johnston 2004]. In most model based analysis of this process, the *bicoid* mRNA is thought to supply proteins at a constant rate in the anterior pole of embryo. However, the available database [Pisarev et al. 2009] and the experimental finding [Surdej and Jacobs-Lorena 1998] support an alternate hypothesis that the stability of mRNA is regulated. Based on these experimental evidences, Liu and Niranjan [2011] proposed three Bicoid concentration computational models in which the maternal *bicoid* mRNA is regulated by being held constant for 2 hours and followed by rapid decay. The uncertainty of such source regulation model is also verified later by Gaussian processing in [Liu and Niranjan 2012].

In this work, two approaches of modelling Bicoid morphogen concentration gradient are compared. The first approach, stochastic chemical reaction system, already used in Liu and Niranjan [2011] and [2012], is used to model the propagation of the diffusion rate of changes using the stochastic modelling in MATLAB software. We then implement another model in agent-based model using FLAME, (Flexible Large-scale Agent-based modelling Environment), to compare the results and the problems faced in both modelling approaches.

The stochastic method provides a way to look at the Bicoid reaction-diffusion model in detail. Observing the details of how Bicoid molecules react with each other, especially the novel mRNA regulation, modellers are able to understand the cellular systems in detail. Stochastic chemical reactions describe the time evolution of the chemically reacting system, in which molecules come in whole numbers and exhibit dynamic behaviour. The basic idea of stochastic simulation was introduced by Gillespie [1977] and Gillespie algorithm is widely used in solving master equation and simulating stochastic process.

The second approach, agent-based modelling (ABM) is another technique inspired from cellular automata methods, quickly becoming the preferred technique among modellers for arguments of accuracy and reliable results. Introduced by Reynolds in 1985, agent-based models have recently become the driving force in various research areas, especially after the advent of more powerful parallel computers. These allow simulations of large populations of agents to be executed in controlled environments, examining the effects of various rules of interactions among the agents. Agent-based models encourage bottom-up, allowing the researchers to focus on the individual elements interacting with each other rather than looking at the complete scenario as a whole. Being stochastic systems, the model patterns can also be defined using differential equations with common examples of its applications found in economic modelling, where mathematical formulas are still being used to prove the behaviour of ideas.

Modelling of complex system behaviour is an emergent science which best demonstrates complex, social behaviour of different communities working together in real world scenarios. Agent-based modeling is a technique which models systems alternatively to the conventional differential equation method modelling. This approach allows a bottom-up procedure where the focus concentrates on the individual interacting units which are given clear defined rules and allowed to simulate. The produced emergent pattern of system behaviour, can then be studied to test and understand the behaviour of complex systems which is otherwise not possible from studying these systems from an outside view. There are various agent based environments introduced which allow modellers to design and test their models. Each of these environments is based on different computational models allowing their usage to vary from their understanding to the computational languages being used to write models. Grimm et al. [2006] discusses a detailed overview of the problems encountered for verifying research work because the tools being used are themselves not being designed on predefined software methodologies.

This paper presents how a model written using stochastic methodology can be converted into an agent-based modelling presenting a comparison between the two techniques. We extend the stochastic model with more details about dynamical behaviour and introduce the agent-based model to set up Bicoid gradient with the realistic mRNA regulation. By comparison between these two models of essentially the same model description, we present a unique comparison in terms of computational details as well as the use from a biological perspective for using the two techniques for understanding complex systems.

## 2. MODELLING METHODOLOGIES

### 2.1. Stochastic Modelling

Closely following [Erban et al. 2007] and [Wu et al. 2007], the stochastic Bicoid protein reaction diffusion system, we implemented a simulation of 100 compartments along the AP axis, each with length h = 5 µm, which is approximately the average size of one nucleus. Refer to Figure 4 for a block diagram.

The three chemical reactions involved in this description are:
- Diffusion of Bicoid proteins along all the compartments:

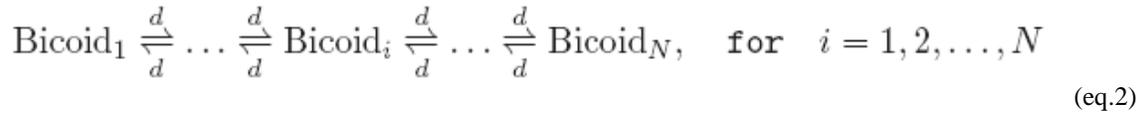

$$\text{Bicoid}_1 \underset{d}{\overset{d}{\rightleftharpoons}} \ldots \underset{d}{\overset{d}{\rightleftharpoons}} \text{Bicoid}_i \underset{d}{\overset{d}{\rightleftharpoons}} \ldots \underset{d}{\overset{d}{\rightleftharpoons}} \text{Bicoid}_N, \quad \texttt{for} \quad i = 1, 2, \ldots, N$$

(eq.2)

- Degradation of Bicoid proteins in all the compartments:

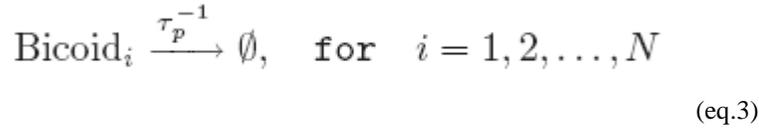

$$\text{Bicoid}_i \xrightarrow{\tau_p^{-1}} \emptyset, \quad \texttt{for} \quad i = 1, 2, \ldots, N$$

(eq.3)

- Translation of *bicoid* mRNA in the anterior pole of embryo (the first compartment).

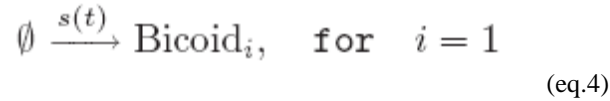

$$\emptyset \xrightarrow{s(t)} \text{Bicoid}_i, \quad \texttt{for} \quad i = 1$$

(eq.4)

where the source regulation function s(t), following [Liu and Niranjan 2011], is given by

$$s(x, t) = s_0 \delta(x) \left( \Theta(t) - \Theta(t - t_0) \right)$$
$$+ s_0 \delta(x) \Theta(t - t_0) \exp\left( -\frac{t - t_0}{\tau_m} \right),$$

(eq.5)

The first of these, Equation 2, describes diffusion between neighbouring sub-volumes, allowed to take place in both directions, at a rate 'd', related to the diffusion constant of a deterministic model by d = D/h2. The second, Equation 3, describes protein degradation, and the final, Equation 4, the source. Translation only takes place in the first bin, for *i = 1*.

Bicoid reaction-diffusion master equation is given by:

$$\frac{\partial}{\partial t}P(\mathbf{n},t) = d\underbrace{\sum_{i=1}^{N-1}[(n_i+1)P(R_i^{\pm 1}\mathbf{n}) - n_i P(\mathbf{n})]}_{\text{Diffusion}: A \to P}$$

$$+ d\underbrace{\sum_{i=2}^{N}[(n_i+1)P(L_i^{\pm 1}\mathbf{n}) - n_i P(\mathbf{n})]}_{\text{Diffusion}: A \leftarrow P}$$

$$+ \tau_p^{-1}\sum_{i=1}^{N}[(n_i+1)P(K_i^{+1}\mathbf{n}) - n_i P(\mathbf{n})]$$

$$+ s(t)P[(K_1^{-1}\mathbf{n}) - P(\mathbf{n})],$$

(eq.6)

where P(**n**, t) is joint probability of state vector **n** = [n1, n2, . . . , ni, . . . , nN] and N = 100. $R_i^{\pm 1}, L_i^{\pm 1}, K_i^{+1}$ and $K_i^{-1}$ are state operators, which are defined by:

$$R_i^{\pm 1}\mathbf{n} = [n_1, n_2, \ldots, n_i + 1, n_{i+1} - 1 \ldots, n_N], \quad i = 1, 2, \ldots, N-1$$
$$L_i^{\pm 1}\mathbf{n} = [n_1, n_2, \ldots, n_{i-1} - 1, n_i + 1 \ldots, n_N], \quad i = 2, 3, \ldots, N$$
$$K_i^{+1}\mathbf{n} = [n_1, n_2, \ldots, n_i + 1, \ldots, n_N], \quad i = 1, 2, \ldots, N$$
$$K_i^{-1}\mathbf{n} = [n_1, n_2, \ldots, n_i - 1, \ldots, n_N], \quad i = 1, 2, \ldots, N$$

The first line in the chemical master equation corresponds to the Bicoid proteins diffusion throughout the A-P axis of the *Drosophila* embryo. The second line describes proteins degradation while the final part is protein synthesis from *bicoid* mRNA. s(t) is mRNA regulation function given by Equation 5.

Our implementation of the Gillespie algorithm for stochastic simulation of the master equation closely follows that of [Erban et al. 2007] and is given in pseudo-code format in Algorithm 1. This process consists of the generation of two random numbers to select the time at which a reaction occurs, and which one that is. The probability that j-th chemical reaction taking place is given by: $a_j/a$, where '*a*' is a total propensity function, computed in step 2 (Algorithm 1). The vector **m** contains the number of molecules along the N = 100 bins while Equations 2 – 4 define a total of *R = 3N – 1* reactions. The propensity functions for the reactions are:

$$\text{Bicoid}_1 \xrightarrow{d} \ldots \xrightarrow{d} \text{Bicoid}_N : \quad a_1 = \sum_{i=1}^{N-1} dm_i$$

(eq.7)

$$\text{Bicoid}_1 \xleftarrow{d} \ldots \xleftarrow{d} \text{Bicoid}_N : \quad a_2 = \sum_{i=2}^{N} dm_i$$

$$\text{Bicoid}_i \xrightarrow{\tau_p^{-1}} \emptyset : \quad a_3 = \sum_{i=1}^{N} \tau_p^{-1} m_i$$

(eq. 8)

The propensity function for the source part is defined by $a_4 = s(t)m_1$ because this reaction occurs in the first bin.

The results for Bicoid stochastic reaction-diffusion in one stochastic simulation realisation based on the Gillespie algorithm *Direct Method* (Algorithm.1) is shown in Figure 7. The model parameters are given as $D = 3$ μm$^2$/s; $t_0 = 144$ min; $\tau_p = 86$ min; $\tau_m = 9$ min. More details for model parameter estimation are discussed in [Liu and Niranjan 2011].

---

**Algorithm 1:** Bicoid reaction-diffusion stochastic simulation

---

**Input**: Model parameters; Final time.
**Output**: Bicoid molecular numbers along 100 compartments: **m**.
**m** = 0; t = 0;
**repeat**
1. Generate two random numbers which are uniformly distributed in (0, 1): r(1) and r(2).

2. Calculate propensity functions of all the reactions: a = a1 + a2 + a3 + a4.

3. Calculate the time when next reaction occurs: $t + \tau$, where $\tau = 1/a \ln(1/r(1))$.

4. Decide which reaction occurs at P $t + \tau$: find $j \in R$ such that: $\sum_{i=1}^{j-1} a_i/a \leq r(2) < \sum_{i=1}^{j} a_i/a$.

5. Update numbers of reactants and products in j-th reaction and set $t \leftarrow t + \tau$.
 **until** *time>final time*;

---

## 2.2. Converting to an Agent-based model

Miller and Page [2007] and Epstein [2007] have favoured agent-based approaches by saying that research should be intensified in this area, to focus into the individual agents rather than the whole systems in equations, realistically allowing individuals to be modelled as agents rather than differential equations. Figure 1 depicts the process involved in writing an agent based model. The model starts with a description about the individual elements in the system which will be represented as agents. The agents are given a set of memory variables, functions and communication protocols which allows them to communicate with each other and the environment. Agents are implemented as separate pieces of code which communicate with each other through communication protocols, also known as messages.

Over the years various platforms have been released for ABM building each using different programming languages and having their own characteristics. Xavier and Foster [2007] and Railsback et al. [2008] have provided a detailed comparison between various platforms by implementing similar models on the different platforms. A comparison of the implementation frameworks is shown in Table I.

To convert any given model to an agent-based model, the modeller first needs to think about the individual agents and their interactions over time. Each of these agents has their own memory and functions as well as messages they send to each other. Figure 2 depicts the structure of an agent communication with another agent.

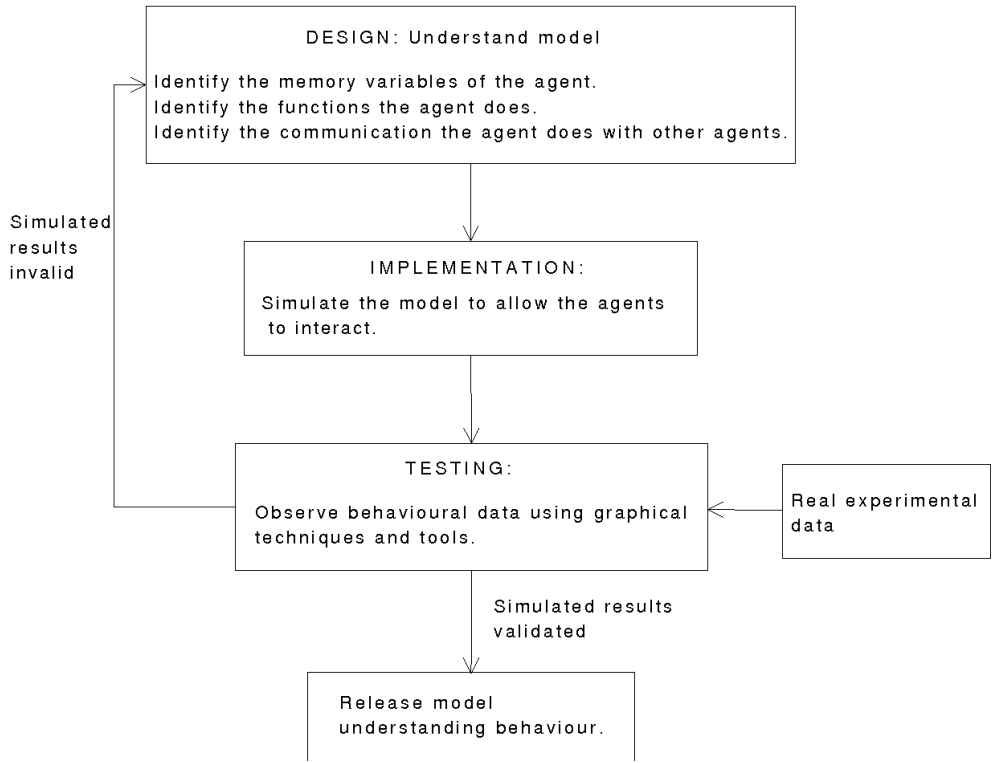

Fig. 1. The order of processes involved when writing an agent-based model. Adapted from [Kiran et al. 2008].

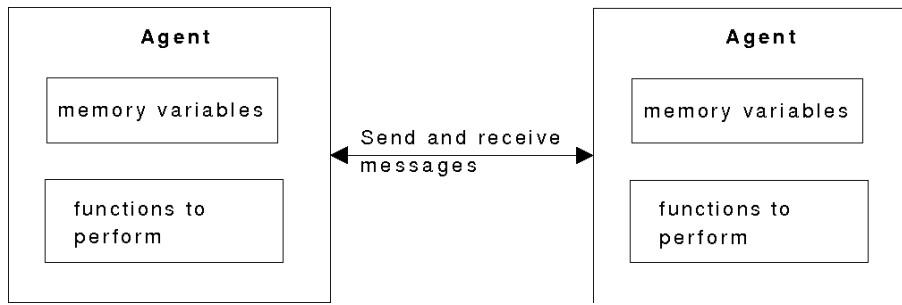

Fig. 2. Structure of a basic agent in the modelling process. Agents can represent any kind of individual like molecules or cells.

Table I. A comparison of commonly used agent-based modelling frameworks *Source:* Adapted from [Kiran et al. 2008] *Note:* Comparison of available platforms.

|  | **SWARM** | **JADE** | **MASON** | **RePast** | **FLAME** |
|---|---|---|---|---|---|
| Software methodology | Programmed in Objective C, and implemented over a nested structure | Uses predefined protocols | Programmed in Java and implemented over a layered structure | Programmed in Java | Programmed in C and designed using X-machine approach |
| Visualisation | 3D | 3D | 3D | 2D | 2D, 3D |
| Parallel or serial | Both. Need to wrap Objective C commands in Java for parallel. | Both | Both | Both | Uses HPC and MPI message for faster communication |
| Examples of models executed | Sugarscape, variety from other disciplines | Virus epidemics, Sugarscape | Virus epidemics, Sugarscape, traffic simulation | Mostly social science projects | Skin grafting, economic modeling |

Agent-based modelling allows each individual molecule to be programmed separately. The level of detail one individual can represent depends on the level of granularity the modeller wants to go into. Figure 3 depicts how SWARM allows nested hierarchies of swarms to be developed where each level can be scheduled with its own scheduler.

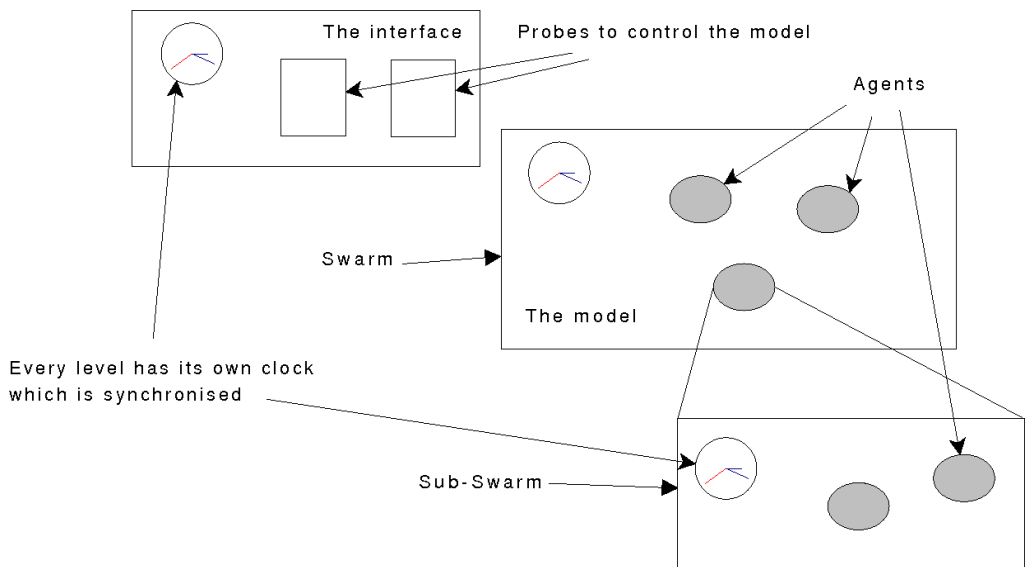

Fig. 3. Nested hierarchy of Swarms or smaller models.

Agents can be designed to represent other sub-swarms which contain their own set of agents and function at different times. Various levels can be introduced in agent based models to encourage details of the workings of the model. The top layer represents the interactions at the surface level. Each of these entities can be connected below to a number of lower layer models receiving and sending changes based on the interactions at the different layers. Each layer can have a different time frame which it covers. For instance the lower layers can function every second and the top layer can work on a per minute basis [Adra et al. 2008]. This is particularly useful when modelling biological models where the models can form a hierarchy of various models together.

Following from the use of the Gillespie algorithm direct method, the embryo cell will be converted into an agent-based model. Figure 4 depicts a structured view of an embryo to understand how a protein diffuses through the length of the embryo. As assumed with the direct method in Gillespie algorithm, the embryo cell can be divided into 100 compartments, with the source sitting in the first compartment. The source will be producing proteins at a certain rate 'r' in this compartment. Depending on another constant rate of diffusion the individual proteins will diffuse into the next compartment to move forward down the length of the embryo depending on the concentration gradients across the membranes. Algorithm 2 describes the steps as mapped to Algorithm 1.

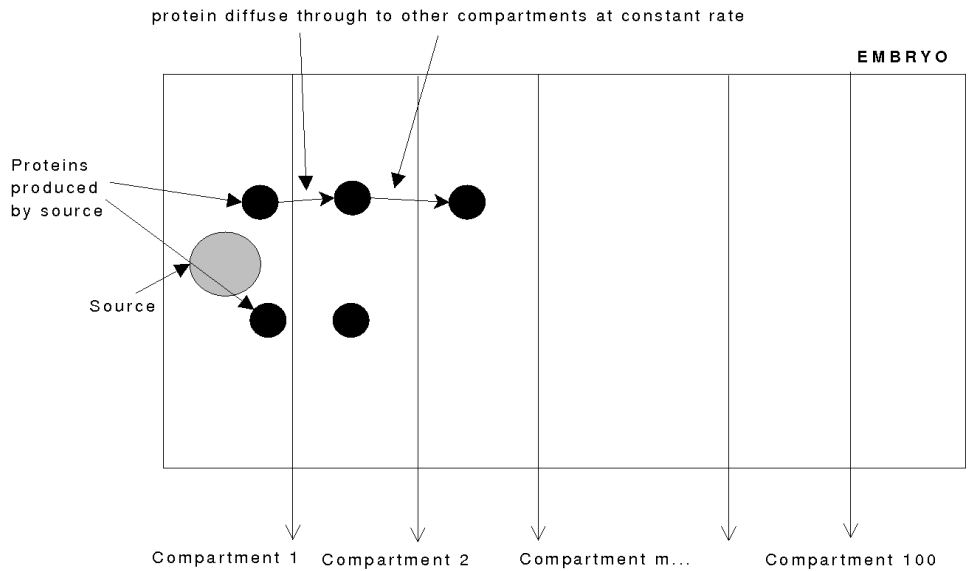

Fig. 4. Movement of proteins within an Drosophila embryo. A structured view.

| **Algorithm 2:** Bicoid reaction-diffusion agent-based simulation |
| --- |
| **Input**: Model parameters (agent memory variable values) at time t=0; Final time of simulation = number of iterations.<br>**Output**: Bicoid molecular numbers along 100 compartments: **m**.<br>**m** = 0; t = 0;<br>**repeat**<br>1. Generate protein production rate as uniformly distributed in (0, 1)<br>2. If source not decayed, calculate probability of producing the protein.<br>3. Decide which reaction occurs allowing molecules to move to the left or right.<br>4. Update numbers of molecules in each compartment.<br>**until** *time>number of iterations*; |

Most modelling techniques involve the same procedures by starting with the problem, decomposing it to simpler subproblems, and trying to solve each subproblem separately. In this manner, multiple problems can be solved simultaneously. Observing the results, can allow the modellers to monitor the problems and gather the information to be analysed later. Based on the description above, any entity or individual which will do any function in the model, can become an agent. For instance, these can be as follows:

- **Source agent.** The source agent is producing new protein agents. The source will also be decaying after a certain time period to reduce its life and eventually disappear or stop producing proteins.
- **Protein agent.** The protein agents are diffusing across the compartment depending on the protein concentration in the compartment it is in.
- **Compartment agent.** To depict the embryo as a whole, a hundred compartment agents can be used to determine the concentration of proteins with in them. This agent can be avoided as the compartments themselves are not doing any function themselves but can be used to account for the result analysis later. Alternatively, one agent representing the environment can also be used. Thus showing that it depends on the modeller's perspective on how he/she writes the model.

Therefore the flow of actions, within an iteration of all the agents involved, can be depicted as in Figure 5. The source, protein and compartment agents perform the functions in the order shown. In addition some of the functions may output certain messages, for instance, the function: Protein posts location, outputs a protein location message which can then be read by other proteins or compartment agents. This message would be the input to the functions for Protein agent to calculate the next move or for the Compartment agent to count how many proteins it has.

### 2.3. Implementation of agent-based model

The above agent-based model was implemented in FLAME (Flexible Large-scale Agent-Based Modelling Environment), is an agent-based modelling environment which enables modellers to write their own models of a variety of complex systems. The simulations written in FLAME have allowed various levels of complexity from modelling molecules to complete communities, by only varying the agent definitions and functions [Coakley et al. 2006]. Formal X-machines are used as the agent architecture, which brought in structure, memory, states and transition functions to the agent. The X-machine agents communicate through messages using the interaction rules specified in the model xml file. These involve posting and reading messages from the message boards (Figure 6).

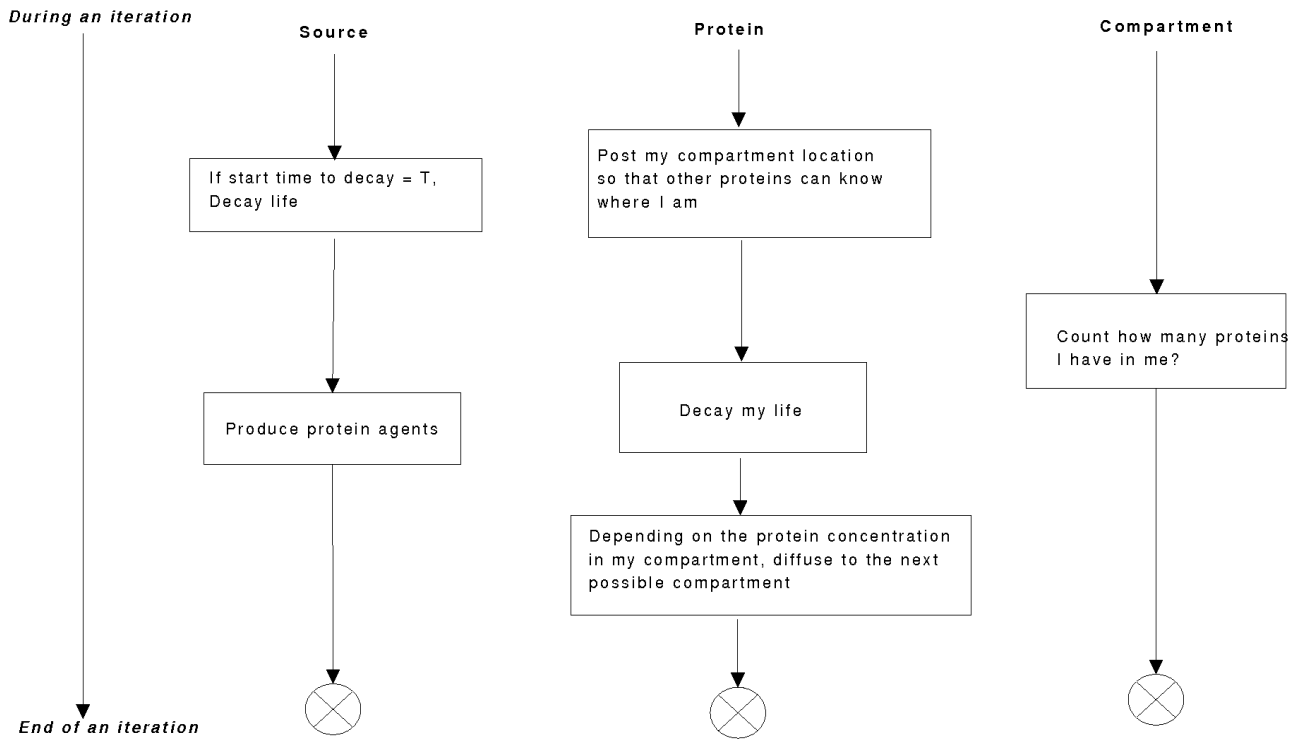

Fig. 5. Agent activities during iteration.

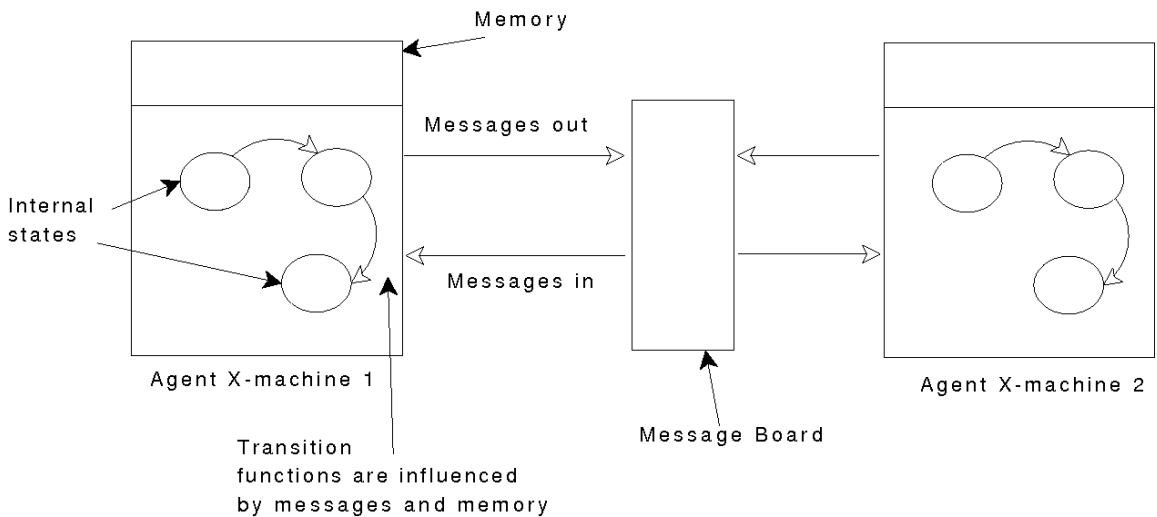

Fig. 6. Two X-machine agent communicating through the message board library. The message board library holds the messages during simulation time step.

The model was executed for 12000 iteration steps assuming that one time step represented one second of the diffusion model (in the stochastic method) and therefore 12000 iterations steps represent two hours of the stochastic equation model.

For starting a simulation, the starting conditions of a model are very important and may influence the model results as time passes. The starting conditions in the agent based model at time *t = 0* were as follows:

- SOURCE DECAY RATE 0.01 - The decay rate of the source.
- SOURCE TIME DECAY 8640 - The time step at which source will start decaying represents 144 minutes of the stochastic simulation.
- SOURCE TIME PRODUCE 50 - The time after which the source will start producing a protein.
- SOURCE PRODUCTION PROB 1.0 - the production probability of a protein
- PROTEIN DECAY RATE 0.01 - the decay rate of the protein.
- COMPARTMENT DIMENSION X 5 - The dimension width of a compartment.
- COMPARTMENT DIMENSION Y 15 - The dimension height of a compartment.
- PROB RIGHT 0.5 - The probability of a protein to move right to the compartment on the right.
- PROB LEFT 0.1 - The probability of a protein to move left to the compartment of the left. The probability to move right was kept higher as this would be more favourable.
- DIE 0.001 - Numerical value to denote when a life goes below this value, kill the agent.

The values for probabilities of source productions rates, protein decay rates, right and left movement probability are conditions which can be changed which each experimental run. These have to change until the ultimate conditions are found. These values were set to above after successive runs of the experiment with other values and these were the closest to the stochastic simulation results.

## 3. RESULTS

To ensure a correct comparison of the two techniques, both models were simulated for identical conditions and data was collected and analysed. Table II represents a comparison of some of the basic simulation details in the two models. The experience in both simulation techniques was compared across a number of factors like simulation time, the memory size needed and tools used.

Analysing the time to actually write the models can be arguable, that it depends on the experience of the programmers. If a programmer who does not have any prior knowledge of agent-based modelling may take more than a month to get accustomed to the manner in which agents have to be defined and communicated with over the platforms. This should also include an installation and learning time for the actual agent-based platforms. For stochastic simulation, MATLAB was used which specialises in mathematical function writing so would be relatively easier to grasp than a different agent-based modeling framework being written for other purposes and the different varieties of these available in different programming languages.

Both models were able to give results which could be analysed in different manners. Agent-based models produce results in terms of timestamp snapshots for the agent conditions at those times. MATLAB was able to produce concentration gradients which could show how the overall system was behaving at different times. This shows one of the major differences in using the two modelling approaches as to what level of detail is required from the models for studying them.

Table II. Simulation comparison between the equation model and the agent-based model *Source:* Programming experience *Note:* Simulation comparison.

| Objective | stochastic model | agent-based model |
|---|---|---|
| **Total simulation time** | 200 min for one realisation. Total time step is around $3 \times 10^6$ (stochastically) | 12000 time steps with 1 second one time step |
| **Actual simulation time** | CPU time: 795.02 seconds | 5 hours |
| **Memory usage** | approx 420 MB | approx 30 GB |
| **Results format produced** | $3.1 * 10^6$ by 100 matrix in MATLAB | 120,000 xml files which are later parsed to produce excel sheets to plot graphs |
| **Actual model writing time** | 1 week. Understanding Gillespie Algorithm and implementation in MATLAB | 1 month, involves understanding the model description and converting to what happens in one iteration |
| **Global values which can easily be changed** | All the decay, production and diffusion rates highlighted in starting conditions | All the decay, production and diffusion rates highlighted in starting conditions |
| **Simulation tool used** | MATLAB 2010b | FLAME serial version run on a MAC laptop |
| **Results were measured** | The results have measured every minute according to all the compartments as shown in Figure 7. In Figure 8, the protein distributed at times 60 min ($9.3 \times 10^5$ iteration step (it)), 100 min ($1.6 \times 10^6$ it), 144 min ($2.2 \times 10^6$ it) 180 min ($2.8 \times 10^6$ it), 200 min ($3.1*10^6$ it) | As number of protein's per time step across compartments, and protein distributions at times 60 min (3600 iteration step (it)), 100 min (6000 it), 144 min (8640 it) 180 min (10800 it), 200 min (12000 it) |
| **Average over runs** | One realisation was taken. The averaged stochastic model is shown by PDE in [Liu and Niranjan 2011] | Model was run 20 times and the average was taken |

The global values can be another deciding factor in how the results fare up in the end. These can be tested through various experimental runs in order to find the optimum conditions for the simulation to give results which match closest to the real data.

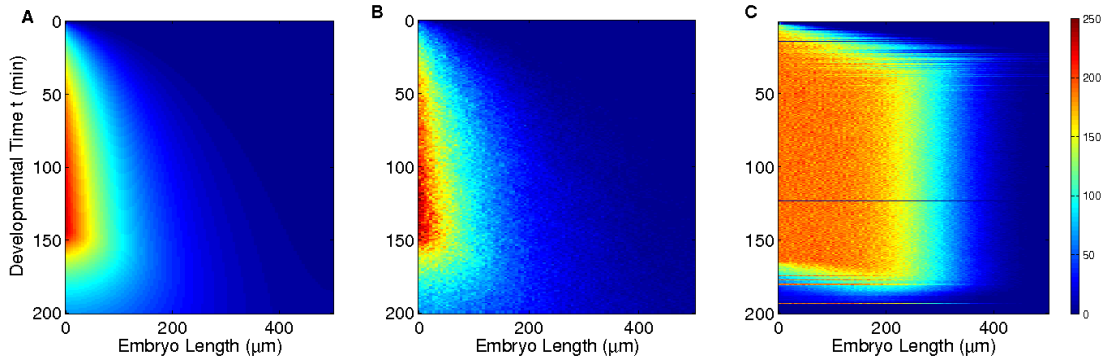

Fig. 7. Bicoid concentration profiles jointly in A-P axis and developmental time. **A** shows deterministic model output which is the average value of the stochastic model. **B** shows one realisation stochastic simulation. **C** shows the result of one agent-based model run.

Agent-based modelling also involves using averages over a number of runs because of the random nature of the agents inherent in the models. Grimm et al. [2006] have argued that averages help reduce the randomness of these models to find the overall behaviour of the model. However this can be done in some situation when overall variables of the system are being measured as in the stochastic simulation. If individual molecule behaviour is being mapped then only single runs of the model can be used to determine how emergence progresses in the system. Therefore an average of 20 run was taken for an overall picture of the system, but in some cases only one simulation run can be considered which makes it necessary for each experimental run to be analysed separately causing data memory issues later.

Figure 7 depicts the intensity plots of the protein distribution across the embryo during the simulation. Figure 7 A shows deterministic model which was reproduced from [Liu and Niranjan 2011]. Figure 7 B and 7 C represents the stochastic and agent-based results respectively. This figure shows a comparison of the molecule concentration in both stochastic and agent-based model. Comparison to the real data the stochastic simulation was able to produce the closest result to the experimental output. However the agent-based model, despite changing the global conditions was not able to come close to the spread of molecules. This is reflected in Figure 8 where the peaks of molecule numbers in the compartments can be compared to the peaks in the stochastic simulation.

Figure 8 represent the distribution along the length of the embryo. The results are shown at identical times (60 min - 3600 iter), (100 min - 6000 iter), (144 min – 8640 iter), (180 min - 10800 iter) and (200 min - 12000 iter).

It is notable that In Figure 8, there is inconsistence of molecule number between the two approaches. Such variance is mainly due to the decay factor in the agent-based model which could not allow the molecules to die out as quickly as in the stochastic simulation. This is because if the decay rate is too high, the protein agents too quickly before reaching the last end of the embryo cell (last compartment) which was needed as one of the basic conditions from the stochastic simulation. Therefore there should be another factor which influences the molecule decay rate in an agent-based model which the stochastic simulation does not pick up.

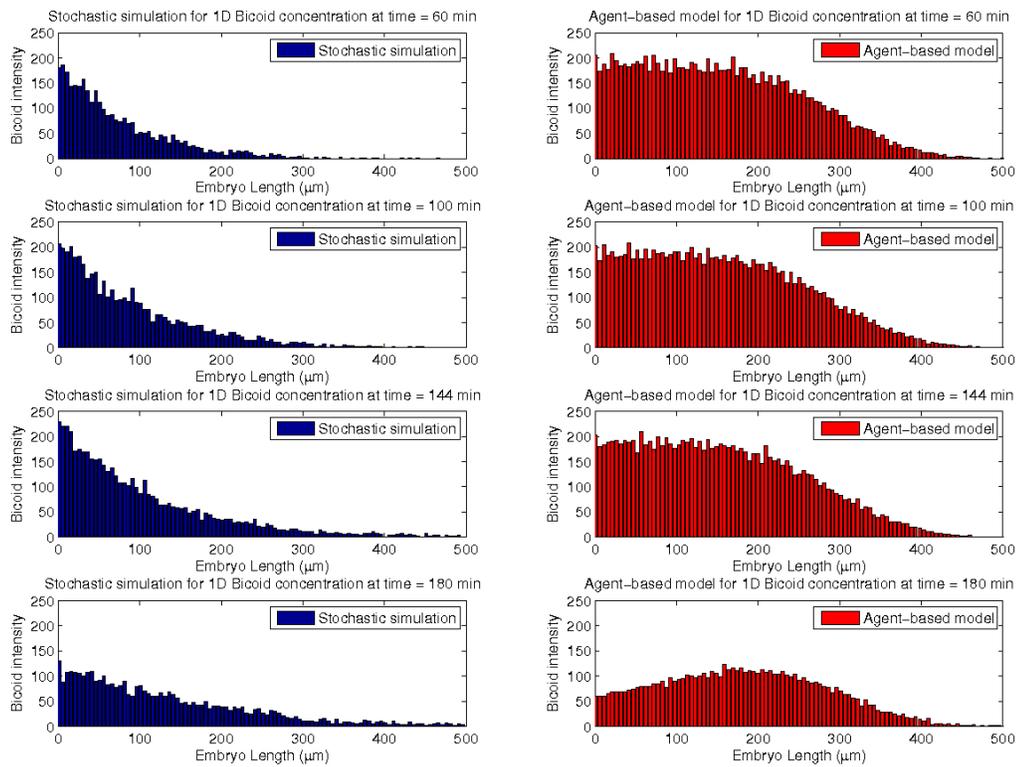

Fig. 8. One realisation stochastic simulation by using Gillespie Algorithm at different time point: 60 (**A**), 100 (**B**), 144 (**C**) and 180 (**D**) min. Blue histograms show the number of Bicoid molecules along anterior and posterior axis in embryo. Red lines show the average amount of molecules which are from deterministic reaction diffusion model [Liu and Niranjan 2011]. Bicoid intensity at 144 min (**(C)**) is the peak stage and it will degrade after this point since the regulation of mRNA. Red histograms show the number of Bicoid molecules along anterior and posterior axis in embryo resulting from average 20 runs of the agent-based model simulation.

Using the same initial conditions from the stochastic model, we were not able to replicate the results in the agent-based simulations (Figure 8). Figure 9 shows the missing data points in the resulting figures when both models used the same initial setting. This led to the agent-based model to be simulated multiple times with different sets of global conditions to find the best set of values which produce results closer to the stochastic model. These values and the ranges they were simulated for are shown in Table III.

Due to the large number of cases used, a minimum square distance was used to calculate the error rate between the results of each of the case with the results.

Table III. Different cases of initial values simulated for the agent-based modelling. The total number of cases simulated for 500 different cases.

| Global value | Range of Values set, interval used | Best Case found (Case 205) |
|---|---|---|
| Source decay rate | [0.01-0.05], 0.1 | 0.03 |
| Protein decay rate | [0.01-0.05], 0.1 | 0.01 |
| Probability of the protein to move right | [0.1-0.5], 0.1 | 0.2 |
| Probability of the protein to move left | [0.1-0.5], 0.1 | 0.3 |
| Source production rate | [0.2-1.0], 0.1 | 0.7 |

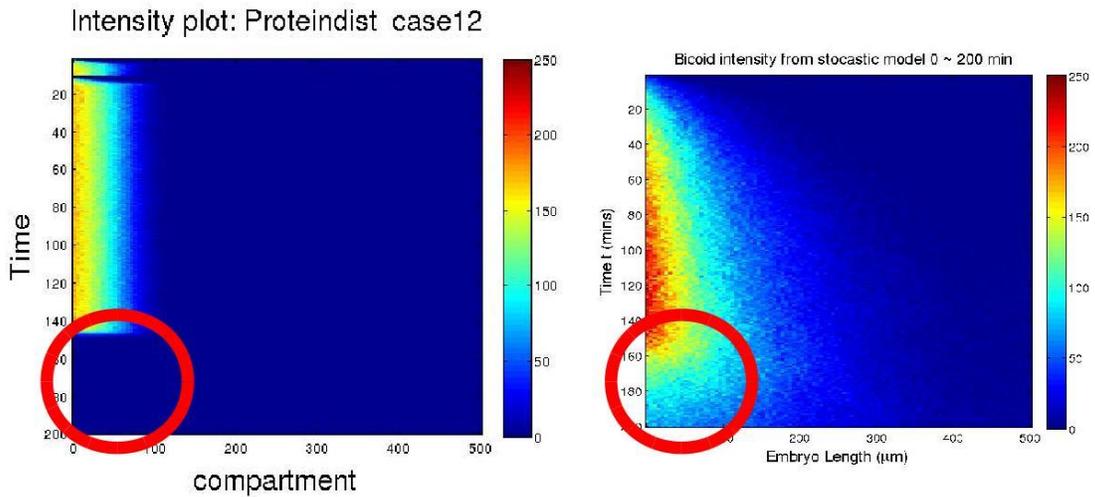
Fig. 9. The agent based modelling simulation result with the stochastic model. The circle shows the missing data points in the agent-based result using the same initial settings in both models.

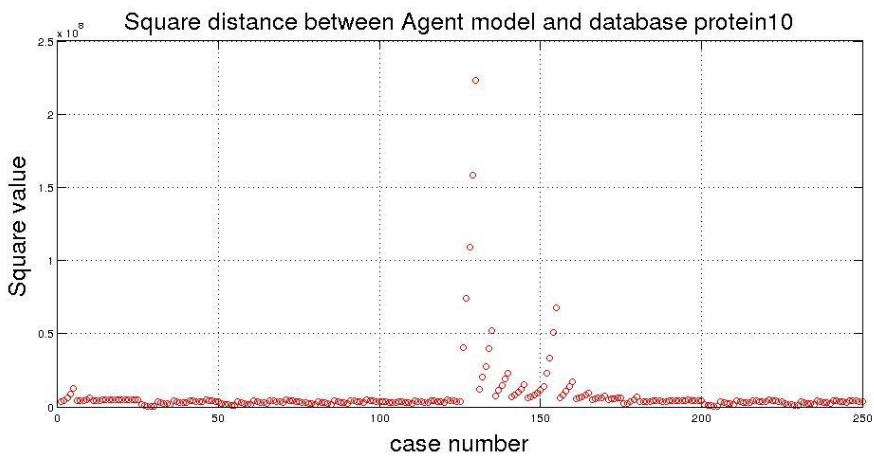
Fig. 10. The Square distance between the agent and the original database results.

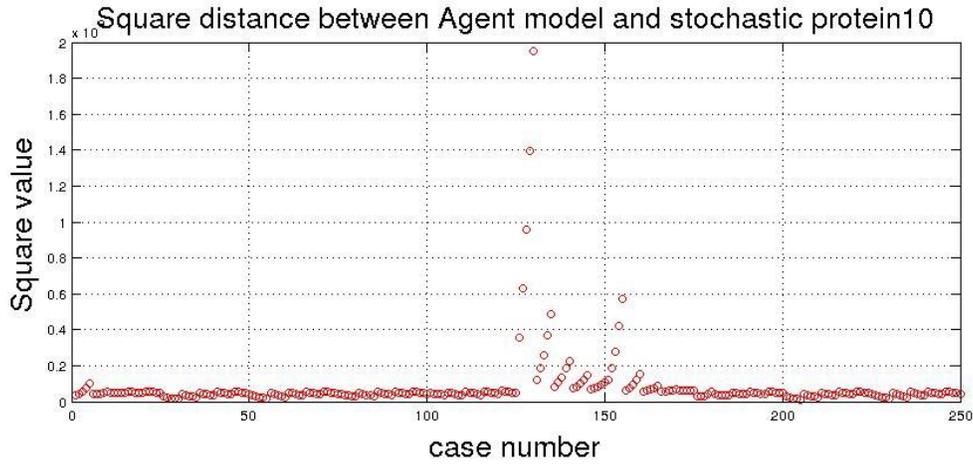

Fig. 11. The Square distance between the agent and the stochastic model.

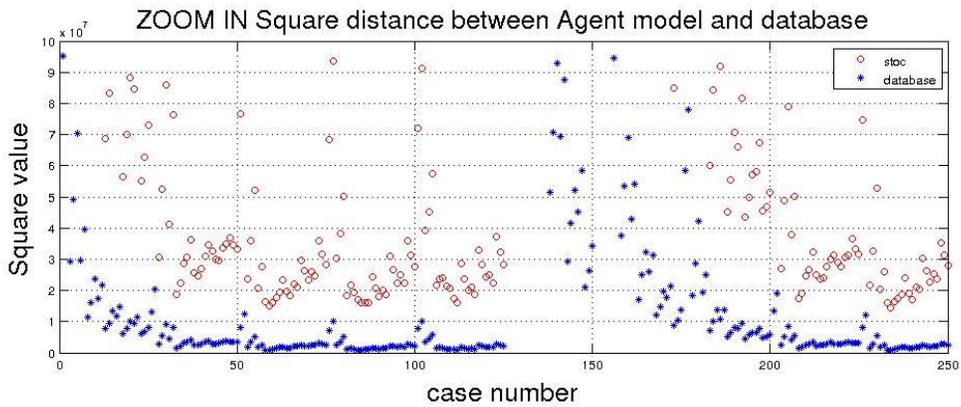

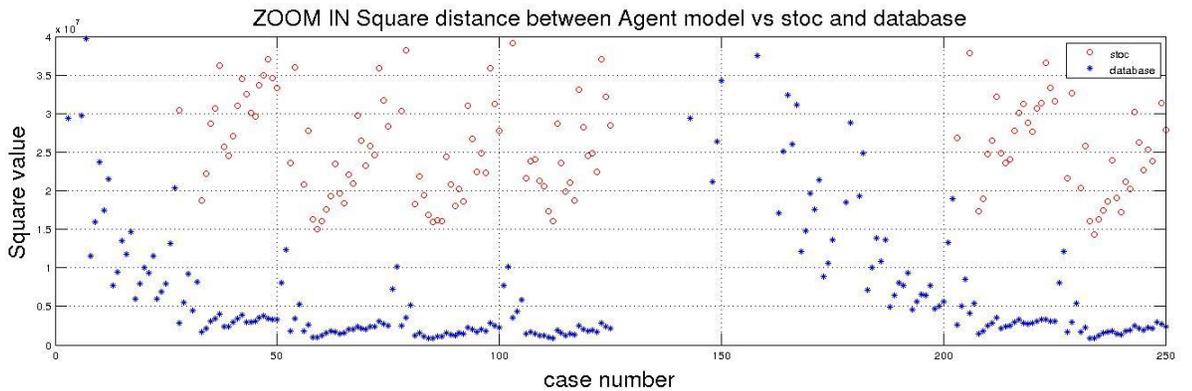

Fig. 12. Zooming in to find the shortest possible error between the simulated cases of results compared to the agent-based results, stochastic and original database results.

Based on the minimum square error method, we were able to find the best possible conditions to use for the agent-based model to duplicate the results of the stochastic model. This was a case selected and shown in Figure 13, using Case 205 with values as defined in Table III.

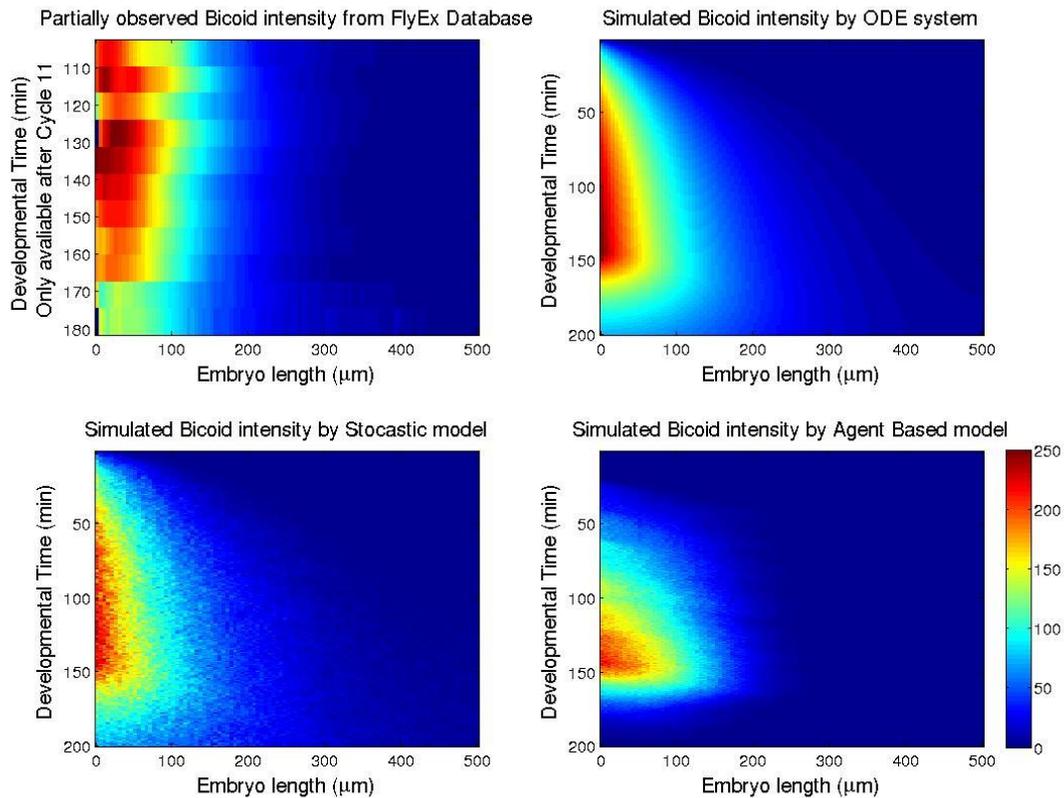

Fig. 12. Zooming in to find the shortest possible error between the simulated cases of results compared to the agent-based results, stochastic and original database results.

## 4. DISCUSSIONS

In the past, various researchers have attempted to compare modelling techniques such as Norling [2007] who presented a comparison of systems dynamics and agent based in a food web evolution. Norling argued that while the systems dynamics approach discovered new insights into the model, agent-based modelling can be used as a tool to further examine the different assumptions made for the system dynamics model to work.

In this experiment, while modelling the Bicoid diffusion as an agent based model as comparison to stochastic simulation, various advantages and disadvantages were witnessed. These can serve as the deciding factors when certain systems have to be analysed.
- **Discovering more details of the model.** This allows modellers to go into the detail for every element's workings, identifying the individual element's memory and their functions which may influence the progress of the model. In equation models, because equations collectively represent agent function as one programming code, modeller is robbed with this opportunity to find new behaviours as a result of this analysis.
- **Global values.** It was observed that global conditions influence the results produced. Global values can be changed dynamically during the course of the model simulation. This can essentially be done in both kinds of models depending on how the models are written.

- **Starting conditions of the model.** Starting conditions can have an effect on the model's results. This can be seen in both kinds of approaches when the simulation results depend on the *t – 1* results.
- **Dynamic inputs to the model.** In an agent-based model, various dynamic agents can be introduced which get activated or influence the progression of results. This can easily programmed by having an agent added which performs certain activities at a certain time steps. This would however be tedious to be programmed in a equation model as complicated nested for-loops may need to be added to the model to allow this. This involves very little changes in an agent-based model.
- **Increasing complexity.** Further complexity can be easily introduced in agent-based models by adding agents and new functions. In equation models, this would require rewriting of the equations and the source code.
- **Directed behaviour.** Agents are autonomous, goal-directed and sociable elements. The decisions they make is based on bounded rationality which means that each agent would have a sphere of influence which allows proximity to be checked before making decisions. In equation models this concept is not present. Here a list is traversed and everything in the list is acted upon in the same manner. In agents, the messages in the sphere of influence may vary allowing agents to display different behaviours depending on where they are located (This is where the concept of emergence comes up). This is particularly useful when modelling realistic biological models.
- **Heterogeneous Populations.** Different agents who differ in memory can be introduced together in the same simulation. This can produce more interesting results as it brings heterogeneity and how agents' internal characteristic can influence the results. This cannot be done in equation models as these assume a homogenous population.
- **Bounded Rationality.** Agents will act depending on their surroundings producing emergent phenomenon. This cannot be programmed in an equation model.
- **Different scenarios during simulations.** Easily different conditions can be introduced to test the model across various conditions. This would not require doing any changes to the agent-based models. Simulations can be stopped half way, conditions can be changed or new agents can be introduced at adhoc and then simulations can be preceded.
- **Large amount of Data produced.** This is a problematic task to analyse the large amounts of data being produced by agent-based models as compared to equation model. Sometimes it is good to find patterns which may not have been thought of previously but this can be a cumbersome task and may require additional intelligent data mining algorithms at a later stage.

## 5. CONCLUSION

The results and experience in both modelling approaches showed that it largely depends on the research questions being attempted when the model for the system is being written. It shows that the biologists need to think about what they are looking for in the model and then choose appropriate techniques accordingly. The best modeling methods may vary in the allowance of what they allow. However there is a huge gap between the disciplines of biologists and computer scientists which is another deciding factor when modelling techniques are chosen. Most biologists are not comfortable with using massive parallel computers for their models or alternate platforms for their work. In the same manner computer scientist are unaware of the needs of the biologists to write platforms which will cater to their needs and what they want to achieve through their systems. Due to this learning curve being quite high for non-computer scientists, research has not progressed when the two disciplines have to be merged.

Our results have shown that modelling techniques are good to study the systems in the larger problem but, we should consider what are the questions being aimed to be answered. If the goal is to find the average concentrations then the differential equations would be best, but if we have queries which go into the individual molecules then agent-based would be better. The results also showed that using the same values for settings in both models will not present the same results of the simulations. This raises concerns for future research is previous models being adapted using newer technologies, showing that this cannot be done easily without losing inherent model properties. Thus models may not be essentially replicated but may be needed to be rewritten for use with new technologies.

Further research work needs to be extended to understand the iterative process between computational modeling and biological data, aiding to selection of the most reliable model based on wet lab experiments. This needs to be extended to find which kinds of applications would suit which kind of modelling techniques in the domains of biology, or others disciplines such as economics or social sciences, raising another case for collaborative research between computer scientists and other disciplines for realizing the full potential of correct and best modelling techniques.